\documentclass[12pt]{iopart}
\usepackage{graphicx}

\begin{document}

\title{Role of inflexible minorities in the evolution of alcohol consumption}

\author{Nuno Crokidakis, Lucas Sigaud}

\address{
Instituto de F\'isica, Universidade Federal Fluminense, Niter\'oi, Rio de Janeiro, Brazil}

\ead{nunocrokidakis@id.uff.br}

\begin{abstract}
\noindent
In this work we study a simple contagion model for drinking behavior evolution, but including the presence of inflexible or zealot agents, i.e., individuals that never change their behavior (never drink or always drink a lot). We analyze the impact of such special agents in the evolution of drinking behavior in the population. Our analytical and numerical results indicate that the presence of only one class of inflexible agents destroys one of the two possible absorbing phases that are observed in the model without such inflexibles. In the presence of the both kinds of inflexible agents simultaneously, there are no absorbing states anymore. Since absorbing states are collective macroscopic states with the presence of only one kind of individuals in the population, we argue that the inclusion of inflexible agents in the population makes the model more realistic. Furthermore, the presence of inflexible agents are similar to the introduction of quenched disorder in the model, and here we observe the suppression of a nonequilibrium phase transition to absorbing states, which had not been reported before.

\end{abstract}

\maketitle

\section{Introduction}
\label{S:1}

A myriad of contagion processes have been studied by epidemic models that go far beyond the scope of infectious diseases~\cite{bailey1975mathematical} and into the very diverse area of social behavior dynamics, such as corruption~\cite{nuno_jorge}, cooperation~\cite{lima2014evolution}, obesity~\cite{ejima2013modeling}, ideological conflicts~\cite{marvel2012encouraging}, fanaticism~\cite{stauffer2007can}, rumor spreading~\cite{daley1964epidemics}, etc. The applications for such models also include general populational behavior, such as rising (and falling) of ancient empires~\cite{gunduz2016dynamics}, violence~\cite{nizamani2014public}, radicalization~\cite{galam2016modeling}, tax evasion dynamics~\cite{brum2017dynamics,ijmpc2022}, legal guns and crimes \cite{nuno_guns} and - of particular interest to this work - addiction.

Regarding addiction, epidemic models have been used previously to describe the populational consumption evolution of different types of substances, from tobacco~\cite{PMID:21696936} and alcohol~\cite{santonja2010alcohol} to heavier drugs such as cocaine~\cite{sanchez2011predicting} and heroine~\cite{heroine2021}. Although addiction is very often seen and treated as an individual's condition, social interaction with people that are more or less users of the respective substance may influence the individual's reaction and consumption, by peer pressure, condemnation or positive reinforcement, ultimately changing their consumption levels. Therefore, it can be treated like a contagion process mediated by the social interactions of individuals with different degrees of addiction. Since alcohol, in particular, is one of the most socially accepted addictive substances for consumption and commerce, as well as one in which social peer pressure can be most effective~\cite{morrislarsen}, it is a prime candidate to be modelled by epidemic models~\cite{galea2009social,gorman2006agent}, specially due to the increase of alcohol consumption not only by social interactions but also spontaneously, which has been documented as a consequence of depression, isolation and even the COVID-19 pandemics~\cite{sullivan2005,clay2020alcohol,narasimha2020complicated,rehm2020alcohol}.

Recently, we addressed the evolution of alcoholism as an epidemic using a compartment contagion model~\cite{nuno_lucas} that subdivided the adult population into three groups, following the alcohol consumption categorization of the World Health Organization~\cite{oms2018}, namely nonconsumers, social (or moderate) consumers and excessive (or risk) consumers. We treated the ensemble as a fully-connected population subdivided into three groups that could influence individuals from one to the other, as well as spontaneously migrate from social to excessive drinkers.

In this work, we study the influence of inflexible individuals in both nonconsumer and excessive consumer groups. Inflexible, or zealot, individuals are the ones that never change their status - in this case, either nonconsumers that never acquire the habit of drinking or excessive drinkers that never diminish their alcohol consumption. This is analogous to introducing disorder in the system, considerably altering its critical behavior. This has been studied before in many contagion modelling of social interactions, such as corruption spreading~\cite{nuno_jorge}, opinion dynamics~\cite{galam1991,martins,celia_vitor} and electoral voting ~\cite{mobilia1}, and we are interested in understanding its impact on the alcohol-consumption model structure of our previous work on the subject~\cite{nuno_lucas}.

Thus, we describe the modelling itself in the next section, along with a brief outline of previous models on alcohol consumption. In Section 3, we present and discuss the results obtained with the new model with inflexibles, considering the presence of either i) inflexible nonconsumers; ii) inflexible excessive consumers; or iii) both types of inflexibles. We follow that with some final remarks and conclusions obtained in this work, and we also include at the end appendixes with the bulk of the mathematical analytical deductions.


\section{Model}

Our model is based on the proposal of Refs. \cite{santonja2010alcohol,galea2009social,gorman2006agent,nuno_lucas,walters2013modelling,nazir2019conformable,huo2018dynamics,mulone2012modeling,sanchez2007drinking,sharma2013drinking,agrawal2018role,adu2017mathematical,huo2012global,khajji2020discrete,huo2017optimal,muthuri2019modeling,ma2015modelling,wang2014optimal,din,rahman} that treat alcohol consumption as a disease that spreads by social interactions. In short, we consider a population of $N$ individuals, which is divided in 3 compartments \cite{nuno_lucas}, namely:

\begin{itemize}
  
\item \textbf{S}: nonconsumer individuals, individuals that have never consumed alcohol or have consumed in the past and quit. In this case, we will call them Susceptible individuals, i.e., susceptible to become drinkers, either again or for the first time;

\item \textbf{M}: nonrisk (or moderate) consumers, individuals with regular low consumption. We will call them Moderated drinkers;

\item \textbf{R}: risk (or excessive) consumers, individuals with regular high consumption. We will call them Risk drinkers;
\end{itemize}

This simple compartmental model was proposed recently \cite{nuno_lucas}, following many other three-compartmental models for alcohol consumption, albeit with structural differences~\cite{santonja2010alcohol,gorman2006agent,walters2013modelling}. Although more complex models incorporated one or more other compartments, such as relapsing individuals~\cite{sanchez2007drinking}, recovered individuals~\cite{sharma2013drinking}, in-treatment and recovered~\cite{khajji2020discrete}, admitting and non-admitting heavy drinkers~\cite{huo2012global} and non-linear incidence rates~\cite{agrawal2018role,adu2017mathematical}, or even external influence on the mobility between compartments (e.g. the effect of mass media campaigns~\cite{huo2017optimal,muthuri2019modeling}, awareness programs~\cite{ma2015modelling}, or hindered interactions between susceptibles and consumers~\cite{wang2014optimal}), the authors focused only on compartments corresponding to the three main alcohol consumption categories, listed above, according to the World Health Organization~\cite{oms2018}, in order to compare with available data~\cite{nuno_lucas}. They incorporated elements from models from both Santonja {\it et al.}~\cite{santonja2010alcohol} (spontaneous transition) and Gorman {\it et al.}~\cite{gorman2006agent} (transition mediated by social interaction) and verified the occurrence of two distinct active-absorbing phase transitions \cite{atman}: one of the phases presents only $S$ individuals, and the other one presents only $R$ individuals. 

Thus, in addition to the interaction rules defined in \cite{nuno_lucas}, we considered here that some agents in the population act as inflexibles or zealots, similar to what has been applied to opinion dynamics models~\cite{nuno_jorge,galam1991,martins,celia_vitor}. Such type of individuals has been given much attention in the past two decades in studies on how radical individuals can impact not only democratic debates~\cite{galam_jacobs,galam2011} and elections~\cite{mobilia1,mobilia2} but also the physical dynamics of phase transitions~\cite{nuno_pmco}. In this case, we consider that we have a certain number $S_{I}$ of inflexibles related to nonconsumer individuals, i.e., individuals that have never consumed alcohol and they will keep this behavior during all their lives. In addition, we also considered a certain number $R_{I}$ of inflexibles related to risk drinkers, i.e., individuals that will always consume alcohol in great quantities. In the following subsections we will elaborate upon the motivations to include such kind of inflexible individuals in the population. Thus, the total number of nonconsumer individuals are given by $S_{total}=S + S_{I}$ and the total number of risk drink individuals are given by $R_{total}=R+R_I$, where $S$ and $R$ denote the noninflexible individuals of each class. Since we are not considering inflexible individuals for the moderated drinkers, we have simply $M_{total}=M$.

\begin{figure}[t]
\begin{center}
\vspace{6mm}
\includegraphics[width=1.0\textwidth,angle=0]{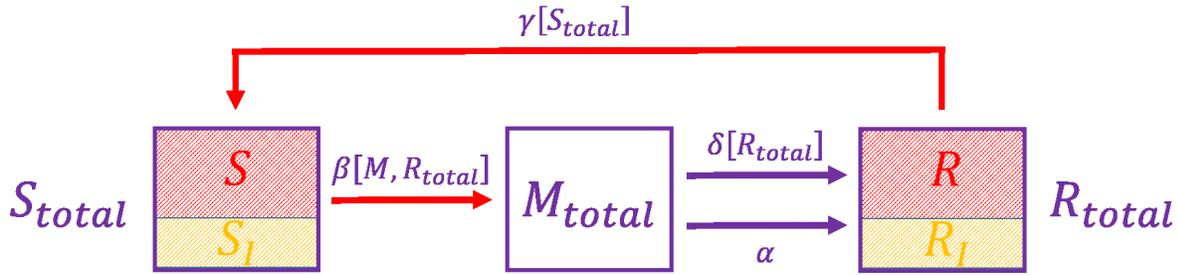}
\end{center}
\caption{(Color online) Schematic representation of the proposed model. Nonconsummers and risk drinkers total populations, respectively $S_{total}$ and $R_{total}$, are subdivided into flexibles (red), $S$ and $R$, and inflexibles (orange), $S_{I}$ and $R_{I}$, respectively. Moderated drinkers are represented by $M_{total}$, and for this class we do not have inflexible individuals. It is important to notice that, while transitions between compartments do not include inflexibles, they contribute to the influence on other compartments, represented by the arrows and their coefficients.}
\label{esquema}
\end{figure}

For this model, the transitions among compartments are the following:

\begin{itemize}
\item $S \stackrel{\beta}{\rightarrow} M$: a noninflexible Susceptible agent (S) becomes a Moderated drinker (M) with probability $\beta$ if he/she is in contact with Moderated (M) or Risk (R and $R_{I}$) drinkers;

\item $M \stackrel{\alpha}{\rightarrow} R$: a Moderated drinker (M) spontaneously becomes a Risk drinker (R) with probability $\alpha$ ;

\item $M \stackrel{\delta}{\rightarrow} R$: a Moderated drinker (M) becomes a Risk drinker (R) with probability $\delta$ if he/she is in contact with Risks drinkers (R and $R_{I}$);

\item $R \stackrel{\gamma}{\rightarrow} S$: a Risk drinker (R) becomes a Susceptible agent (S) with probability $\gamma$ if he/she is in contact with Susceptible individuals (S and $S_{I}$));  

\end{itemize}
A schematic representation of such transitions is shown in Figure \ref{esquema}. All parameters are rates, i.e., probabilities per unit time. As usual in models of drinking behavior evolution \cite{gorman2006agent,sanchez2007drinking,din}, the time unit is given in days. Thus, the parameters $\beta, \alpha, \delta$ and $\gamma$ are given in units of day$^{-1}$.

In this work we consider population homogeneous mixing, i.e., a fully-connected population of $N$ individuals. However, since we are interested in the critical phenomena that occur in the model, we will work in the infinite population limit $N\to\infty$, since phase transitions only make sense in infinite-size systems \cite{marro2005nonequilibrium,hinrichsen2000non,stanley_book}. Thus, instead of populations $S, M$ and $R$, we will consider the populations densities, that will be defined in the next section.

Notice from the rules above that inflexible individuals cannot change compartments, since their opinions/states are frozen in time. However, they can persuade other noninflexible individuals to change compartments, for example a inflexible Susceptible agent $S_{I}$ can persuade a noninflexible Risk drinker $R$ to stop drinking and change to state $S$ with probability $\gamma$.

As discussed in \cite{nuno_lucas}, the rules above are governed by transition probabilities. $\beta$ represents an ``infection'' probability, i.e., the probability that a consumer ($M$ or $R$) individual turns a nonconsumer one ($S$) into a drinker. We considered just one probability $\beta$ for the $S\to M$ transition, whether the susceptible individuals interact with moderate or risk drinkers. One can also consider two distinct probabilities, say $\beta_1$ and $\beta_2$, to consider a more general scenario where a non-drinker may have a higher propensity to starting to drink when interacting with a moderate drinker, and he/she may be put off by interacting with a heavy drinker. However, to avoid an extra parameter in the model, we considered for simplicity $\beta_1=\beta_2=\beta$. The Risk drinkers $R$ or $R_I$ can also ``infect'' the Moderated $M$ agents and turn them into noninflexible Risk drinkers $R$, which occurs with probability $\delta$. This transition $M\to R$ can also occur spontaneously, with probability $\alpha$, if a given agent increase his/her alcohol consumption without social interaction. As stated in the introduction, above, the increase of alcohol consumption has been documented to occur under stressful circumstances (like the COVID-19 pandemic~\cite{clay2020alcohol}) or clinical depression~\cite{sullivan2005}, regardless of social interaction with Risk drinkers. Finally, the probability $\gamma$ represents the infection probability that turns noninflexible Risk drinkers $R$ into noninflexible Susceptible agents $S$. In this case, it can represent the pressure of social contacts (family, friends, etc) over individuals that drink excessively. For simplicity, we did not take into account transitions from Risk (R) to Moderate (M), assuming that, as a rule, once an individual reaches a behavior of excessive consumption of alcohol, contact with Moderate drinkers does not imply on a tendency to lower one's consumption - meanwhile, it is assumed that contacts that do not drink at all are able to exert a higher pressure on them to quit drinking. On the other hand, in our model~\cite{nuno_lucas}, as well as many others~\cite{santonja2010alcohol,gorman2006agent,walters2013modelling,sharma2013drinking,agrawal2018role,adu2017mathematical,huo2012global,khajji2020discrete}, the influence of non-drinkers on risk drinkers is assumed to be much more present and effective than on moderate drinkers.

From now on, we will only consider the densities (relative proportions) of individuals in each compartment, i.e., we will deal only with the subpopulations in relation to the total population. Thus, we will not take into account births and deaths, meaning that the population does not vary with time. As discussed in refs. \cite{nuno_lucas,mulone2012modeling,sanchez2007drinking}, one other way of looking at this approximation is to consider only the adult population as relevant to our modeling, and assume that the number of new individuals coming of age correspond to the number of deaths.

In the next sections we consider three distinct cases, according to inflexibility: (i) there are only $S_{I}$ inflexible individuals in the population ($R_{I}=0$); (ii) there are only $R_{I}$ inflexible individuals in the population ($S_{I}=0$); (iii) both inflexible individuals $S_{I}$ and $R_{I}$ are present in the population.


\section{Results}

\subsection{Presence of $R_I$ inflexible individuals}

\qquad In this section we will consider only $R_I$ inflexible individuals in the population, i.e., there are some individuals that will always consume alcohol in large quantities. It is documented that many risk drinkers are not open to change their habits or even admit their level of alcohol consumption due to many reasons, such as social stigma~\cite{gomberg}, mistrust in governments and/or treatment centers~\cite{kolosova}, personal (or family) denial~\cite{spirito}, etc. It is estimated that, in the United States, only about 1.3\% of almost 19 million people above 12 years old with a substance addiction disorder admit their condition and seek treatment~\cite{nsduh2019}, and these numbers may have gotten even worse during the COVID-19 pandemics~\cite{rehm2020alcohol}. This issue is so relevant that some models subdivide the risk consumers compartment into admitters and deniers~\cite{huo2012global}, and we incorporate it here in the form of risk drinkers that cannot be influenced by nonconsumers. Thus, we are not considering in this section the presence of nonconsumer inflexibles, i.e., we have $S_{I}=0$.

At mean-field level, one assumes a population of infinite size ($N\to\infty$) and ignore any random fluctuations \cite{mobilia3}. Hence, the densities $s(t)=S(t)/N, m(t)=M(t)/N$ and $r(t)=R(t)/N$ are defined at each time step $t$ and are treated as continuous variables. In addition, we have a fixed density of inflexible risk drinkers $r_I = R_I/N$, which does not depend on $t$. For this case, the normalization condition can be written as 
\begin{equation} \label{eq1}
s(t)+m(t)+r(t)+r_I = 1 ~, 
\end{equation}
\noindent
valid at each time step $t$. Based on the microscopic rules defined in section 2, the densities obey the following rate equations (see Appendix A for details):
\begin{eqnarray} \label{eq2}
\frac{ds(t)}{dt} & = & -\beta\,s(t)\,m(t) - \beta\,s(t)\,[r(t)+r_I] + \gamma\,s(t)\,r(t) ~, \\ \label{eq3}
\frac{dm(t)}{dt} & = & \beta\,s(t)\,m(t) + \beta\,s(t)\,[r(t)+r_I] - \delta\,m(t)\,[r(t)+r_I] - \alpha\,m(t) ~, \\ \label{eq4}
\frac{dr(t)}{dt} & = & \alpha\,m(t) + \delta\,m(t)\,[r(t)+r_I] - \gamma\,s(t)\,r(t) ~.
\end{eqnarray}

\begin{figure}[t]
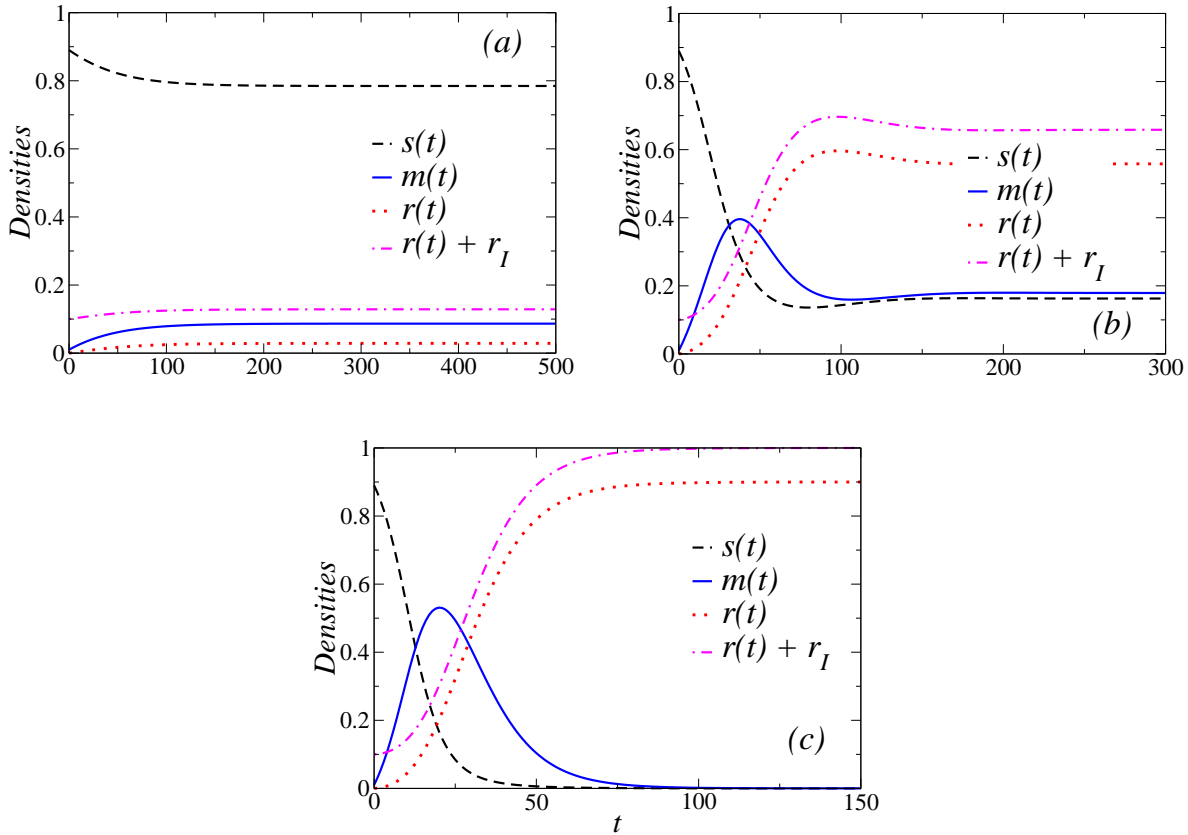

\begin{center}
\vspace{6mm}
\includegraphics[width=0.48\textwidth,angle=0]{figure2a.eps}
\hspace{0.3cm}
\includegraphics[width=0.48\textwidth,angle=0]{figure2b.eps}
\\
\vspace{0.5cm}
\includegraphics[width=0.48\textwidth,angle=0]{figure2c.eps}
\end{center}
\caption{(Color online) Time evolution of the three densities of agents $s(t)$, $m(t)$ and $r(t)$, and the total density of risk drinkers $r+r_I$ for $r_{I}=0.10$, based on the numerical integration of Eqs. (\ref{eq2}) - (\ref{eq4}). The fixed parameters are $\alpha=0.03, \delta=0.07$ and $\gamma=0.15$, and we varied $\beta$: (a) $\beta=0.02$, (b) $\beta=0.10$, (c) $\beta=0.20$  (all parameters are given in units of day$^{-1}$).}
\label{fig1}
\end{figure}

One can start analyzing the time evolution of the compartments $s, m$ and $r$. We numerically integrated Eqs. (\ref{eq2}), (\ref{eq3}) and (\ref{eq4}) in order to analyze the effects of the variation of the model’s parameters. As initial conditions, we considered $m(0)=0.01, r(0)=0.0$ and $s(0)=1-m(0)-r_{I}$ \footnote{As the initial conditions do not affect the stationary values of $s, m$ and $r$, we did not considered other initial conditions.}. We will consider as illustration the case $r_{I}=0.10$, i.e., $10\%$ of the population is composed by inflexible risk drinkers. For simplicity, we fixed $\alpha=0.03, \delta=0.07$ and $\gamma=0.15$, for comparison with the case $r_{I}=0$ \cite{nuno_lucas}, and we considered some values of the infection probability $\beta$. From Fig. \ref{fig1} we can see that the system evolves to stationary states for sufficient long times. For the case with $\beta=0.02$ (Fig. \ref{fig1} (a)), the population does not fall in an absorbing state as in the case $r_I=0$ \cite{nuno_lucas}, even for a very small value of $\beta$. For $\beta=0.10$ (Fig. \ref{fig1} (b)) we observe the coexistence of the subpopulations $s, m, r$ and $r_I$. Finally, for the case $\beta=0.20$ we can see that, for sufficient long times, the system achieves an absorbing state with $s=m=0$ and $r+r_{I}=1$. This state is absorbing since there are only risk drinkers in the population ($R$ and $R_I$). In such a case, the dynamics becomes frozen since no transitions will occur anymore \cite{marro2005nonequilibrium,hinrichsen2000non}. These results will be discussed in more detail below.

Now we can analyze the stationary properties of the model in the presence of inflexible risk drinkers $r_I$. We denote the stationary densities as $s=s(t\to\infty), m=m(t\to\infty)$ and $r=r(t\to\infty)$.
Some details of the analytical calculations are exhibited in Appendix A. There we can find two possible solutions for the density of noninflexible nonconsumers $s$: a trivial solution $s=0$, and another solution $s\neq 0$. If the solution $s=0$ is valid, we can also find analytically that $m=0$ (see Appendix A). From the normalization condition of Eq. (\ref{eq1}), we obtain $r=1-r_{I}$. This solution represents an absorbing state similar to the one observed in the case $s_{I}=r_{I}=0$ \cite{nuno_lucas}, i.e., we have the absence of $S$ and $M$ populations, leaving only risk drinkers in the population, with inflexible $r_I$, and noninflexible $r$, risk drinkers. The first stationary solution for the case $r_I\neq 0$ is then $(s,m,r)=(0,0,1-r_{I})$. The dynamics becomes frozen since risk drinkers cannot leave their compartment in the absence of susceptibles and moderate drinkers \cite{nuno_lucas}.

Let us consider now the case $s\neq 0$. We can see in Appendix A that $m=0$ is not a solution if $r_{I}\neq 0$. Thus, the second absorbing solution observed for $r_{I}=0$, namely $(s,m,r)=(1,0,0)$ \cite{nuno_lucas}, is destroyed in the presence of inflexible risk drinkers $r_I$. Despite the theoretical interest of physicists in absorbing phase transitions \cite{marro2005nonequilibrium,hinrichsen2000non}, from the practical point of view an absorbing phase with no risk drinkers is not realistic. Thus, the presence of $R_I$ inflexible individuals makes the model more realistic, despite the presence of the other absorbing state $(s,m,r)=(0,0,1-r_{I})$. 

\begin{figure}[t]
\begin{center}
\vspace{6mm}
\includegraphics[width=0.6\textwidth,angle=0]{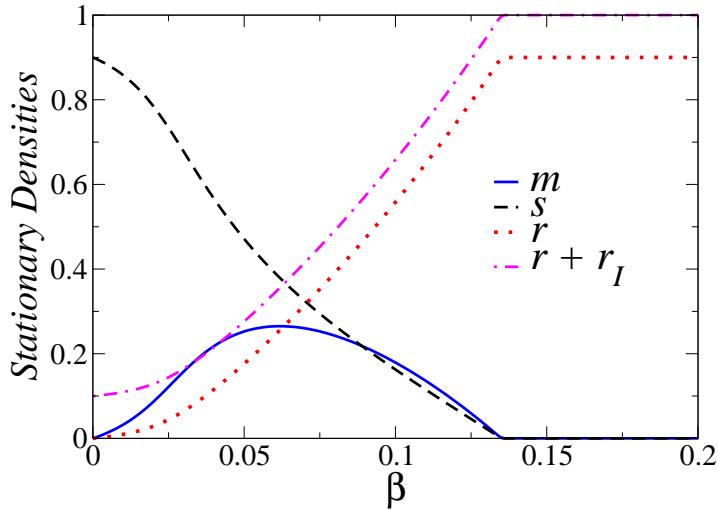}
\end{center}
\caption{(Color online) Stationary densities $m, s, r$ and the total density of risk drinkers $r+r_I$ as functions of $\beta$ for $r_I=0.10$ and $s_{I}=0.00$. The results were obtained by the numerical integration of Eqs. (\ref{eq2}) - (\ref{eq4}). The fixed parameters are $\alpha=0.03$, $\gamma=0.15$ and $\delta=0.07$ (all parameters are given in units of day$^{-1}$).}
\label{fig2}
\end{figure}

An illustration of the stationary states of the model is exhibited in Fig. \ref{fig2}, where we show the stationary densities as functions of $\beta$ for fixed $\alpha=0.03$, $\gamma=0.15$ and $\delta=0.07$. For this figure, we consider $r_I=0.10$, i.e., $10\%$ of the population is formed by inflexible risk drinkers. The results were obtained by the numerical integration of Eqs. (\ref{eq2}) - (\ref{eq4}). We can see that for small values of $\beta$ there is no absorbing phase  as in the case $r_I=0$ ($s=1, m=r=0$ for finite $\beta$) \cite{nuno_lucas}. Even for $\beta=0.0$ there is a coexistence among nonconsumers $s$ and inflexible risk drinkers $r_I$. However, for larger values of $\beta$ ($\beta>\approx 0.13$ in Fig. \ref{fig2}) we observe an absorbing phase $s=m=0$ and $r+r_I=1$, as obtained analytically (see Appendix A). For intermediate values of $\beta$ we observe the coexistence of $s, m, r$ and $r_I$ subpopulations. For the considered parameters, this region is $0 < \beta < \approx 0.13$ in Fig. \ref{fig2}.

Finally, since we are interested in the stationary states of the model, it is easy to observe that the parameters that lead to the increase of the final (stationary) value or $r = r(t\to\infty)$ are $\delta$, $\alpha$ and $\beta$. Meanwhile, if $\gamma$ is decreased $r$ also increases. Regarding the behavior of $r$ with $\beta$, it can be seen directly from Fig. \ref{fig1} that the drinking outbreak (the $m$ peak) increases considerably with the increase of $\beta$, and therefore also the final value of $r$. Eq. (\ref{eqx1}) also gives one a clue: although we have no closed expression for $m$, this equation shows that, by either increasing $\beta$ or decreasing $\gamma$, $r$ is increased. Regarding the final stationary states, the initial values of the three populations don't matter. On the other hand, the increase of social programs or alcoholism-treating campaigns can lead to an effective increase on the $\gamma$ parameter, which works as a lowering of $r$ and, therefore, a higher value of $s$ (in this case, coming from recovery after being classified as risk consumers)~\cite{walters2013modelling,khajji2020discrete}.

In equilibrium systems like magnetic models, the presence of disorder usually destroys phase transitions \cite{belanger,nuno_nobre}. Here, we observe a similar effect in a nonequilibrium system. In this case, the frozen states of the inflexible agents work in the model as the introduction of quenched disorder. As in magnetic systems, one can expect that disorder can induce or suppress a phase transition, as was also observed in the kinetic exchange opinion model in the presence of inflexibles \cite{celia_vitor}. However, the presence of disorder in models with absorbing states does not lead to the destruction of active-absorbing phase transitions, at least in low-dimensional systems \cite{Barghathi1,Barghathi2}. At mean-field level, to the best of our knowledge, the suppression of nonequilibrium phase transitions to absorbing states had not been previously observed. Thus, the model makes a contribution to the study of the impact of quenched disorder in nonequilibrium phase transitions.


\subsection{Presence of $S_{I}$ inflexible individuals}

\qquad In this section we will consider a more commonly encountered scenario, where there are only $S_{I}$ inflexible individuals in the population, i.e., there are some individuals that have never consumed alcohol and they will keep this behavior during all their lives. Such individuals can be representative, for example, of one of the many religions and/or places that forbid alcohol consumption. However, even outside communities where alcohol consumption is forbidden, persons that absolutely refrain from drinking are becoming more common in many regions of the world~\cite{oms2018,holmes2019}, and recent studies show the influence that teetotaller young adults can have in their social environments~\cite{conroy2013}. Thus, in order to analyze the sole influence of inflexible nonconsummers in the model, we are not considering in this section the presence of inflexible risk drinkers, i.e., we have $R_I=0$.

As in the previous subsection, we will work with population densities. In this case, we have a fixed density of inflexible nonconsumers $s_I = S_I/N$, that does not depend on $t$. The normalization condition can then be written as 
\begin{equation} \label{eq7}
s(t)+m(t)+r(t)+s_I = 1 ~.
\end{equation}

Thus, based on the microscopic rules defined in section 2, one can write the rate equations that describe the time evolution of the densities,  
\begin{eqnarray} \label{eq8}
\frac{ds(t)}{dt} & = & -\beta\,s(t)\,m(t) - \beta\,s(t)\,r(t) + \gamma\,[s(t)+s_I]\,r(t) ~, \\ \label{eq9}
\frac{dm(t)}{dt} & = & \beta\,s(t)\,m(t) + \beta\,s(t)\,r(t) - \delta\,m(t)\,r(t) - \alpha\,m(t) ~, \\ \label{eq10}
\frac{dr(t)}{dt} & = & \alpha\,m(t) + \delta\,m(t)\,r(t) - \gamma\,[s(t)+s_I]\,r(t) ~.
\end{eqnarray}  

\begin{figure}[t]
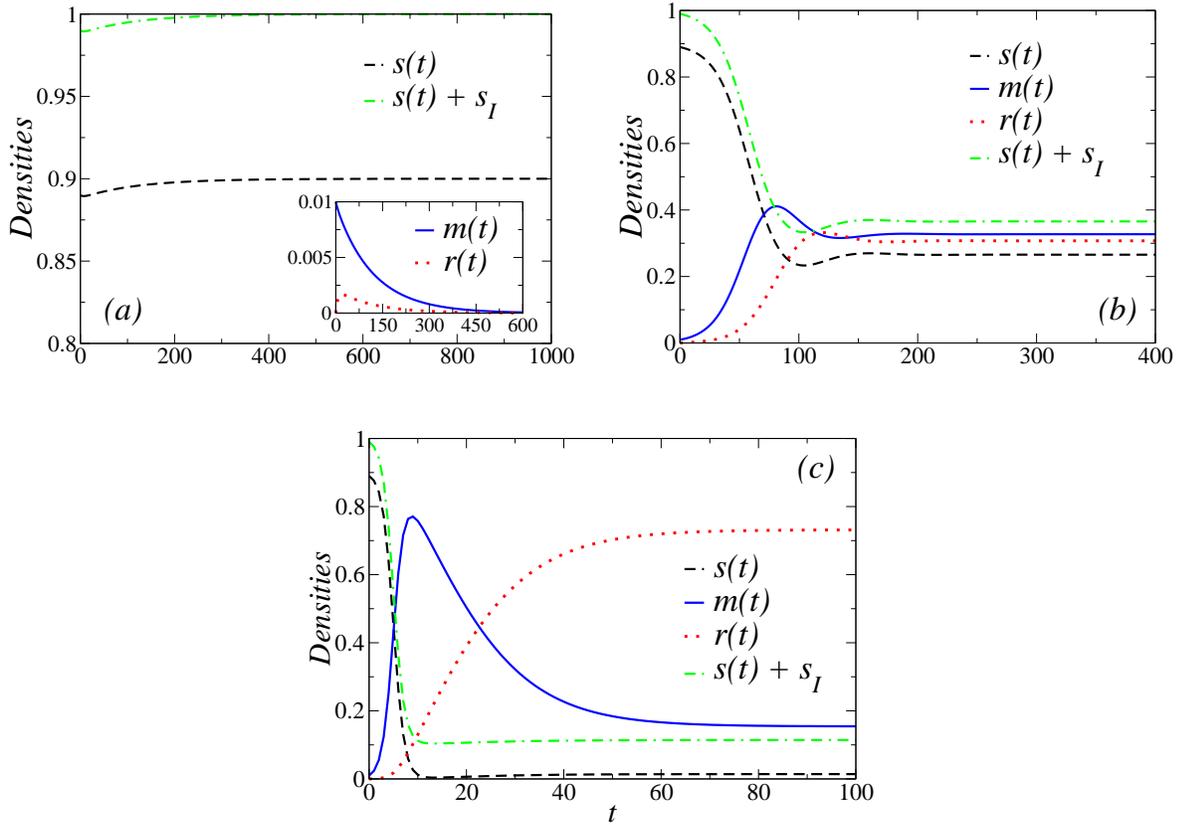

\begin{center}
\vspace{6mm}
\includegraphics[width=0.48\textwidth,angle=0]{figure4a.eps}
\hspace{0.3cm}
\includegraphics[width=0.47\textwidth,angle=0]{figure4b.eps}
\\
\vspace{0.5cm}
\includegraphics[width=0.48\textwidth,angle=0]{figure4c.eps}
\end{center}
\caption{(Color online) Time evolution of the three densities of agents $s(t)$, $m(t)$ and $r(t)$, and the total density of nonconsumers $s+s_I$ for $s_{I}=0.10$, based on the numerical integration of Eqs. (\ref{eq8}) - (\ref{eq10}). The fixed parameters are $\alpha=0.03, \delta=0.07$ and $\gamma=0.15$, and we varied $\beta$: (a) $\beta=0.02$, (b) $\beta=0.10$, (c) $\beta=1.00$ (all parameters are given in units of day$^{-1}$).}
\label{fig3}
\end{figure}

Again, one can start by analyzing the time evolution of the compartments $s, m$ and $r$. We numerically integrated Eqs. (\ref{eq8}), (\ref{eq9}) and (\ref{eq10}) in order to analyze the effects of the variation of the model’s parameters. As initial conditions, we considered $m(0)=0.01$, $r(0)=0.0$ and $s(0)=1-m(0)-s_{I}$. We will consider, as a sample case, $s_{I}=0.10$, i.e., $10\%$ of the population is composed by inflexible nonconsumers. For simplicity, we fixed $\alpha=0.03, \delta=0.07$ and $\gamma=0.15$, for comparison with the case $s_{I}=0$ \cite{nuno_lucas}, and we considered some values of the infection probability $\beta$. From Fig. \ref{fig3} we can see that the system evolves to stationary states for sufficiently long times. For the case with $\beta=0.02$ (Fig. \ref{fig3} (a)), the population falls in an absorbing state with $m=r=0$ and $s+s_{I}=1$, as in the case $s_{I}=0$ \cite{nuno_lucas}. As in the previous section, this is an absorbing state since the susceptible individuals cannot leave their compartment in the absence of moderated and risk drinkers. For $\beta=0.10$ (Fig. \ref{fig3} (b)) we observe the coexistence of subpopulations $s, m, r$ and $s_{I}$. Finally, for the case $\beta=1.00$ (Fig. \ref{fig3} (c)) we can see that, for sufficiently long times, the system does not achieve another absorbing state as in the case $s_{I}=0$ \cite{nuno_lucas}. These results will be discussed in more detail below.

Let us analyze the stationary states of the model. Some details of the calculations are given in Appendix B, in which we can find a trivial solution for the stationary density of noninflexible nonconsumers $r$, i.e., $r=0$. If the solution $r=0$ is valid, we can also find, analytically, $m=0$ (see Appendix B). From the normalization condition (Eq. \ref{eq7}), we obtain $s=1-s_{I}$. This solution represents an absorbing state similar to the one observed in the case $s_{I}=r_{I}=0$ \cite{nuno_lucas}, i.e., we have the absence of $M$ and $R$ populations, and there are only susceptible nonconsumers in the population, combining inflexible individuals $s_{I}$ and noninflexible ones $s$. The first starionary solution for the case $s_{I}\neq 0$ is then $(s,m,r)=(1-s_{I},0,0)$.

\begin{figure}[t]
\begin{center}
\vspace{6mm}
\includegraphics[width=0.6\textwidth,angle=0]{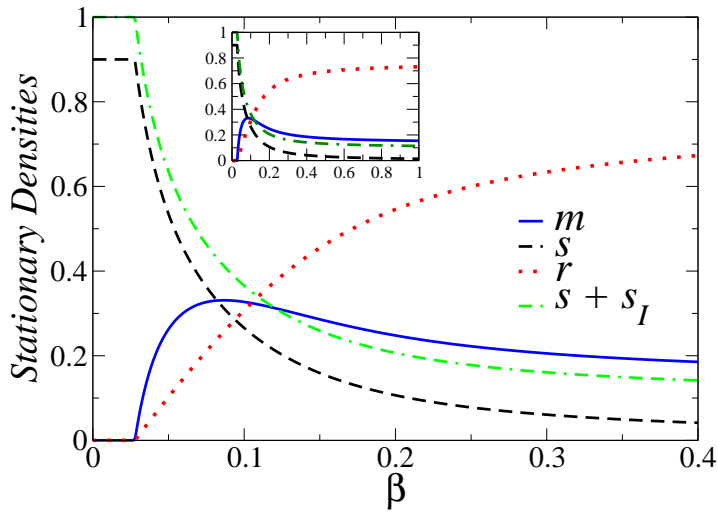}
\end{center}
\caption{(Color online) Stationary densities $m, s, r$ and the total density of nonconsumers $s+s_{I}$ as functions of $\beta$ for $s_{I}=0.10$ and $r_I=0.00$. The results were obtained by the numerical integration of Eqs. (\ref{eq8}) - (\ref{eq10}). The fixed parameters are $\alpha=0.03$, $\gamma=0.15$ and $\delta=0.07$ (all parameters are given in units of day$^{-1}$).}
\label{fig4}
\end{figure}

Let us consider the case $r\neq 0$. We can see from the calculations of Appendix B that $m=0$ is not a solution if $s_{I}\neq 0$. Thus, the second absorbing solution observed for $s_{I}=0$, namely $(s,m,r)=(0,0,1)$ \cite{nuno_lucas}, is destroyed in the presence of nonconsumer inflexible individuals. As discussed in the previous subsection, an absorbing state with the absence of one of the subpopulations is not realistic. In the present case, it is an absorbing phase with no nonconsumers $s$. Thus, the presence of $S_{I}$ inflexible individuals makes the model more realistic, despite the presence of the other absorbing state $(s,m,r)=(1-s_{I},0,0)$. 

An illustration of the stationary states of the model is exhibited in Fig. \ref{fig4}, where we show the stationary densities as functions of $\beta$ for fixed $\alpha=0.03$, $\gamma=0.15$ and $\delta=0.07$. For this figure, we consider $s_{I}=0.10$, i.e., $10\%$ of the population is formed by inflexible nonconsumers. The results were obtained by the numerical integration of Eqs. (\ref{eq8}) - (\ref{eq10}). We can see that for small values of $\beta$ we observe an absorbing phase where $m=r=0$ and $s+s_{I}=1$. However, for larger values of $\beta$ we do not observe an absorbing phase as was observed for $s_{I}=0$ \cite{nuno_lucas}. For intermediate and large values of $\beta$ we observe the coexistence of $s, m, r$ and $s_{I}$ subpopulations. The inset of Fig. \ref{fig4} shows that even for $\beta=1.0$ there is no absorbing phase. For the considered parameters, the values of the densities of nonconsumers is $s\approx 0.015$ and moderate drinkers is $m\approx 0.154$ for $\beta=1.0$.


\subsection{Presence of $S_{I}$ and $R_{I}$ inflexible individuals}

\qquad In this section we will consider both $S_{I}$ and $R_{I}$ inflexible individuals in the population. This describes the coexistence of both inflexible nonconsumers and inflexible risk drinkers, and we will analyze here the influence on the overall population that these two radical groups exert. This is the more general case, with two distinct sources of disorder in the model.

As before, we will work with population densities. In this case, we have fixed densities of inflexible risk drinkers $r_I = R_I/N$ and inflexible nonconsumers $s_{I}=S_{I}/N$, both not depending on $t$. For this case, the normalization condition can be written as 
\begin{equation} \label{eq13}
s(t)+m(t)+r(t)+s_I+r_I = 1 ~.
\end{equation}

Based on the microscopic rules defined in section 2, one can write the rate equations that describe the time evolution of the densities,  
\begin{eqnarray} \label{eq14}
\frac{ds(t)}{dt} & = & -\beta\,s(t)\,m(t) - \beta\,s(t)\,[r(t)+r_I] + \gamma\,[s(t)+s_I]\,r(t) ~, \\ \label{eq15}
\frac{dm(t)}{dt} & = & \beta\,s(t)\,m(t) + \beta\,s(t)\,[r(t)+r_I] - \delta\,m(t)\,[r(t)+r_I] - \alpha\,m(t) ~, \\ \label{eq16}
\frac{dr(t)}{dt} & = & \alpha\,m(t) + \delta\,m(t)\,[r(t)+r_I] - \gamma\,[s(t)+s_I]\,r(t) ~.
\end{eqnarray}

\begin{figure}[t]
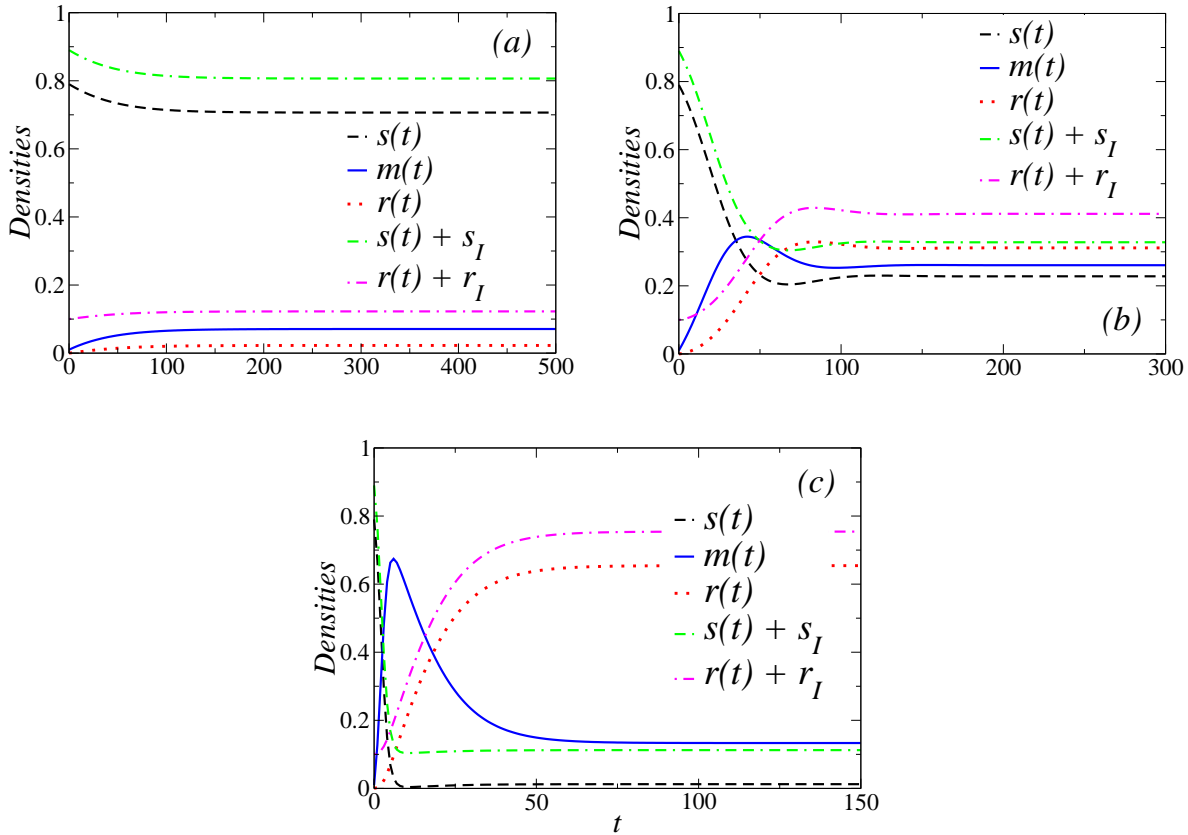

\begin{center}
\vspace{6mm}
\includegraphics[width=0.48\textwidth,angle=0]{figure6a.eps}
\hspace{0.3cm}
\includegraphics[width=0.48\textwidth,angle=0]{figure6b.eps}
\\
\vspace{0.5cm}
\includegraphics[width=0.48\textwidth,angle=0]{figure6c.eps}
\end{center}
\caption{(Color online) Time evolution of the three densities of agents $s(t)$, $m(t)$ and $r(t)$, and the total densities of nonconsumers $s+s_I$ and of risk drinkers $r+r_I$, for $s_{I}=r_I=0.10$, based on the numerical integration of Eqs. (\ref{eq14}) - (\ref{eq16}). The fixed parameters are $\alpha=0.03, \delta=0.07$ and $\gamma=0.15$, and we varied $\beta$: (a) $\beta=0.02$, (b) $\beta=0.10$, (c) $\beta=1.00$ (all parameters are given in units of day$^{-1}$).}
\label{fig5}
\end{figure}

One can start by analyzing the time evolution of compartments $s, m$ and $r$. We numerically integrated Eqs. (\ref{eq14}), (\ref{eq15}) and (\ref{eq16}). As initial conditions, we considered $m(0)=0.01, r(0)=0.0$ and $s(0)=1-m(0)-r_{I}-s_{I}$. We will consider as an example the case $s_{I}=r_{I}=0.10$, i.e.,  $10\%$ of the population is formed by inflexible nonconsumers and $10\%$ of inflexible risk drinkers. As in the previous subsections, we fixed $\alpha=0.03, \delta=0.07$ and $\gamma=0.15$, and we considered some values of the infection probability $\beta$. From Fig. \ref{fig5}, one can readily see that for all shown values of $\beta$ ($\beta = 0.02$ (a), $\beta = 0.10$ (b), and $\beta = 1.00$ (c)) there are no absorbing states whatsoever, i.e., there are no trivial solutions for the equilibrium states of the model. On the contrary, the presence of inflexibles are responsible for maintaining all compartments with occupation values different from zero, promoting the social interactions that keep noninflexible individuals changing their drinking status. As should be expected, for low values of the transmission coefficient $\beta$ (Fig. \ref{fig5} (a)) we have a predominance of nonconsumers in the equilibrium state; on the other hand, for high values of $\beta$ the situation is reversed (Fig. \ref{fig5} (c)), with risk drinkers attaining the prominent role, once a transient phase of moderate consumers majority has passed - actually, the noninflexible nonconsumer population rapidly decreases to near zero and the moderate consumer population tends to almost the same value as the $s + s_{I}$ compartment. Intermediate values of $\beta$ (Fig. \ref{fig5} (b)) lead to a more homogeneous mixing of the three consumers group, which is prone to be a more accurate description of most communities where no alcohol-consumption ban exists~\cite{oms2018}.

\begin{figure}[t]
\begin{center}
\vspace{6mm}
\includegraphics[width=0.6\textwidth,angle=0]{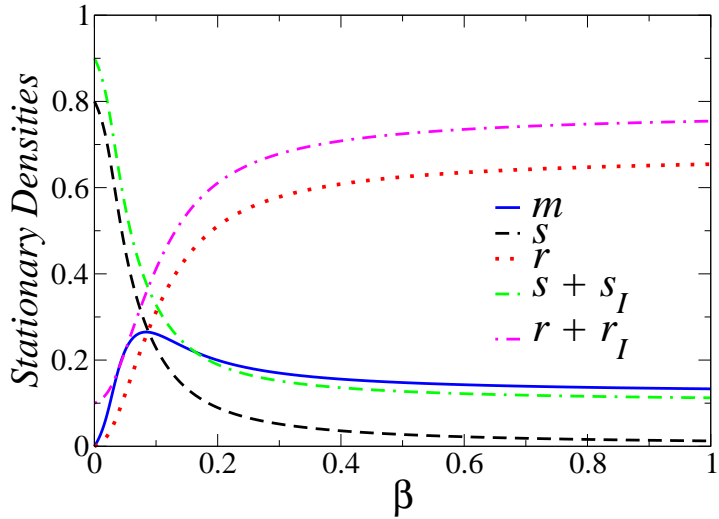}
\end{center}
\caption{(Color online) Stationary densities $m, s, r$ and the total densities of nonconsumers $s+s_{I}$ and risk drinkers $r+r_I$ as functions of $\beta$ for $s_{I}=0.10$ and $r_I=0.10$. The results were obtained by the numerical integration of Eqs. (\ref{eq14}) - (\ref{eq16}). The fixed parameters are $\alpha=0.03$, $\gamma=0.15$ and $\delta=0.07$ (all parameters are given in units of day$^{-1}$).}
\label{fig6}
\end{figure}

An illustration of the stationary states of the model is exhibited in Fig. \ref{fig6}, where we show the stationary densities as functions of $\beta$ for fixed $\alpha=0.03$, $\gamma=0.15$ and $\delta=0.07$. For this figure, we consider $s_{I}=r_{I}=0.10$, i.e., $20\%$ of the population is formed by inflexible agents, $10\%$ nonconsumers and $10\%$ risk drinkers.

The results of Fig. \ref{fig6} corroborate the ones described above for Fig. \ref{fig5}. One can see that only for small values of $\beta$ a majority of nonconsumers are present for the studied conditions, but as $\beta$ increases a predominance of risk drinkers quickly supersedes the other subpopulations. Curiously, no stationary solution with $s = m = r$ can be found in the model, which was the same case for the model without the presence of inflexibles ($s_{I} = r_{I} = 0$)~\cite{nuno_lucas}.

It should also be noted that the increase of the transmission coefficient $\beta$ value leads to a fast predominance of risk drinkers due to a combination of factors: as $R$ increases, more individuals migrate from $S$ to $M$, since $\beta$ depends directly on the $R$ subpopulation; also, the increase of the number of $M$ individuals leads to a spontaneous migration to the $R$ compartment via the $\alpha$ coefficient - since this mechanism is not available for any other transition pathway, the predominance of risk drinkers is present for a much wider range of $\beta$ values than nonconsumers, even though they present, in this study, the same amount of inflexible individuals.

The presence of the two kinds of inflexible agents, $S_{I}$ and $R_{I}$, leads to a system where there are no absorbing transitions anymore. In such a case, as discussed before, the model becomes more realistic, since collective macroscopic states with the absence of at least one of the subpopulations (nonconsumers, moderate drinkers or risk drinkers) are not realistic from the practical point of view.


\section{Final remarks}   

In this work we study a model that constitutes an expansion of a previously studied simplified contagion model for alcoholism consumption~\cite{nuno_lucas}, where the population was subdivided into three compartments, namely nonconsumers, moderate consumers and risk drinkers~\cite{oms2018}.  In this work, we include a fraction of inflexible nonconsumers and/or risk drinkers, and study the subsequent effects on the stationary behavior of their subpopulations.

The first major difference observed between the previously studied simplified model~\cite{nuno_lucas} and the newer, more complex, one presented here is the destruction of one (when one group of inflexible individuals was included) or both (when both inflexible groups are present) absorbing states. This is analog to other equilibrium systems such as magnetic models, where the presence of disorder usually destroys phase transitions \cite{belanger,nuno_nobre}. The introduction of inflexible agents works as the inclusion of quenched disorder in the model, which can suppress a phase transition. However, in this model we observe the suppression of a nonequilibrium phase transition to absorbing states, which had not been previously observed in studies of quenched disorder in nonequilibrium phase transitions.

Furthermore, the absence of previously observed absorbing states shows the continual influence of inflexible - or zealot - groups. The presence of only, for example, radical nonconsumers intermixed with the population rapidly leads the whole group to become nonconsummers over time; likewise, the converse could be observed with the presence of only radical risk drinkers. On the other hand, the presence of both radical groups mingled with the noninflexible individuals leads to final stationary states where all groups still have representatives - i.e., no absorbing states are observed. The relative predominance (or not) of an individual group is directly linked to the transmission coefficient values, representing the power of persuasion between the different compartment groups. This also represents a much more realistic approach to describing worldwide alcohol consumption, since - with perhaps the exception of some communities where alcohol consumption is a criminal offense - nonconsumers, moderate drinkers and risk drinkers alike are present in the society.

It is important to discuss about the model on finite systems of size $N$. Considering a finite population, the model is a finite Markov chain with absorbing states, $R + R_I = N$ for the case where $S_I=0$, and $S + S_I = N$ when $R_I=0$. In these cases, by the general properties of finite absorbing Markov chains, the model is guaranteed to end up in an absorbing state \cite{vankampen}. For finite populations, the finite-size effects are such that in the absorbing phase the final configuration is reached in a characteristic time that scales logarithmically with the system size, while, in the active phase, finite-size fluctuations take the system to a frozen configuration in a characteristic time that grows exponentially with the system size \cite{saeedian}. In this sense, the active states (coexistence of subpopulations) we observed are not stable for finite populations \cite{mobilia3,assaf}. However, in this work we are interested in the critical phenomena that occur in the model. Systems exhibiting a phase transition to an absorbing state possess (for appropriate values of the control parameter) an active (nonabsorbing) stationary state in the infinite-size limit. But if the system size is finite, the process ends up in the absorbing state \cite{oliveira_quasi1,oliveira_quasi2}. As pointed explicitly in \cite{oliveira_quasi2}, \textit{"...the only true stationary state for a finite system is the absorbing one"}. Thus, for the study of the phase transitions, we need to consider the infinite-size limit $N\to\infty$  \cite{marro2005nonequilibrium,hinrichsen2000non,stanley_book}. This justifies the mean-field rate equations we considered for the present study.

As future extensions, the model can be considered in regular lattices, as well as complex networks. In lattices or networks, in addition to the present dynamics of the model studied here, we can consider diluted structures, i.e., lattices/networks with empty sites. In such a case, we can consider mobility of the agents through the network, that can lead to rich phenomenology \cite{nuno_pmco2}. For example, by dividing the modelled population into sectors, the influence of different environments for the increase (or decrease) of drinking behavior can be taken into account, such as college campi~\cite{manthey2008}. In addition, other kinds of special agents like contrarians \cite{galam_cont,galam_cont2,nuno_joao,stauffer_jorge} should also be considered. Also, we can consider a more realistic modeling, similar to references \cite{gutkin,chou}. These models consider cognitive and learning mechanisms, and thus treat addiction in a more fundamental level. For example, in \cite{gutkin} the authors present a neurocomputational model that combines a set of neural circuits at the molecular, cellular, and system levels and accounts for several neurobiological and behavioral processes leading to nicotine addiction. A similar approach can be considered for alcohol consumption. Also, another work considered a dynamical systems model to describe the neurobiology of drug addiction that incorporates the psychiatric concepts \cite{chou}. Such kind of descriptions can be of interest for future studies in dynamics of drinking behavior.

Finally, age-related social dynamics of the interactions can also be incorporated in the model. It has been shown that age difference can be incorporated in social contagion models by triggering different social interactions related to drinking behavior~\cite{guo2018}, making the model a more realistic approach to real social interactions. Not only that, generational conflicts has been shown to also influence heavy drugs intake, such as heroin~\cite{liu2019}, and radicalization behavior~\cite{chuang2018}, which can lead to the growth of zealotry. Therefore, it is an interesting candidate for modification of this model, by dividing our compartmental structure into two different one (youngsters and elders), but with social interactions between them.


\section*{Acknowledgments}

The authors thank Serge Galam for some suggestions. Financial support from the Brazilian scientific funding agencies CNPq (Grants 310893/2020-8 and 311019/2017-0) and FAPERJ (Grant 203.217/2017) is also acknowledged. Finally, we also acknowledge thoughtful remarks by anonymous referees which significantly improved the text.


\appendix

\section{Analytical considerations: Model in the presence of $R_I$ inflexible individuals}

Before discussing the stationary solutions of Eqs. (\ref{eq2}) - (\ref{eq4}), let us discuss how to obtain a rate equation from the master Equation. Following \cite{mario_book,satulovsky}, let us consider a lattice of $N$ sites which can be either occupied by a susceptible individual, a moderate drinker or a risk drinker. The state of the system is represented by $\sigma = (\sigma_{1},\sigma_{2},...,\sigma_{N})$, where $\sigma_i = 0, 1$ or $2$ according to whether the site $i$ is occupied by a susceptible individual, a moderate drinker or a risk drinker. Let $P(\sigma, t)$ be the probability of state $\sigma$ at time $t$ and let $\omega_{i}(\sigma)$ be the probability per unit time of a cyclic permutation of variable $\sigma_i$. That is, if $\sigma_i = 0, 1$ or $2$, then $\omega_{i}$ is the transition probability to $\sigma_i = 1, 2$ or $0$, respectively \cite{mario_book,satulovsky}. The evolution of  $P(\sigma, t)$  is governed by the master equation \cite{mario_book},
\begin{equation} \label{n1}
\frac{d}{dt}\,P(\sigma,t) = \sum_{i=1}^{N} [\omega_{i}(\sigma^{i})\,P(\sigma^{i},t) - \omega_{i}(\sigma)\,P(\sigma,t)] ~
\end{equation}
where the state $\sigma^{i}$ can be obtained from the state $\sigma=(\sigma_{1},\sigma_{2},...,\sigma_{i},...,\sigma_{N})$ by an anticyclic permutation of the variable $\sigma_{i}$ \cite{satulovsky}.

Considering the susceptible population as an example, for a given lattice site $i$ the following transitions are possible (see the rules in section 2):
\begin{itemize}
\item $0 + 1 \stackrel{\beta/z}{\rightarrow} 1 + 1$

\item $0 + 2 \stackrel{\beta/z}{\rightarrow} 1 + 2$

\item $0 + 3 \stackrel{\beta/z}{\rightarrow} 1 + 3$

\item $2 + 0 \stackrel{\gamma/z}{\rightarrow} 0 + 0$

\end{itemize}
where $z$ is the lattice coordination number (number of neighbors of site $i$), and we included a fourth state, $\alpha=3$, to represent the inflexible risk drinkers $r_I$.

Considering the above possible transitions, one can write the following expression for the transition probability per unit time \cite{mario_book,satulovsky}:
\begin{eqnarray}\nonumber
\omega_{i}(\sigma_{i}) & = & \frac{\beta}{z}\,\delta(\sigma_i,0)\,\sum_{\epsilon=1}^{z}[\delta(\sigma_{i+\epsilon},1)+\delta(\sigma_{i+\epsilon},2)+\delta(\sigma_{i+\epsilon},3)] \\  \label{n2}
&& + \frac{\gamma}{z}\,\delta(\sigma_i,2)\,\sum_{\epsilon=1}^{z}\delta(\sigma_{i+\epsilon},0)
\end{eqnarray}
where $\delta(x,y)$ is the Kronecker delta and the summation is over the nearest neighbors of site $i$.

The mean value of any function $f(\sigma)$ can be obtained from
\begin{equation}  \label{n3}
\langle f(\sigma)\rangle = \sum_{\sigma}\,f(\sigma)P(\sigma,t)    
\end{equation}
Considering the master equation (\ref{n1}) and the expression for the transition probability per unit time (\ref{n2}), one can write the time evolution of a general function $f(\sigma)$ as  \cite{mario_book,satulovsky}
\begin{equation}\label{n4}
\frac{d}{dt}\,\langle f(\sigma)\rangle  =  \sum_{i=1}^{N}  \langle [f(^{i}\sigma) - f(\sigma)]\,\omega_{i}(\sigma) \rangle 
\end{equation}
where the state denoted by $^{i}\sigma$ is obtained from $\sigma$ by a cyclic permutation of the variable $\sigma_{i}$ \cite{satulovsky}.

For our model, the functions $f(\sigma)$ of interest are given by the site correlations, namely
\begin{eqnarray}\label{n5}
P_{i}(\alpha) & = & \langle \delta(\sigma_{i},\alpha) \rangle \\ \label{n6}
P_{ij}(\alpha\,\beta) & = & \langle \delta(\sigma_{i},\alpha)\,\delta(\sigma_{j},\beta) \rangle 
\end{eqnarray}

In Eq. (\ref{n5}), we have the single site correlation $P_{i}(\alpha)$, that is the probability of a random chosen lattice site $i$ is in state $\alpha=0,1,2,4$, where $0$ represents the susceptible individuals, $1$ represents the moderate drinkers, $2$ represents the risk drinkers and $3$ denotes the inflexible risk drinkers. The another equation, (\ref{n6}), denotes the two-site correlations $P_{ij}(\alpha,\beta)$ \cite{mario_book,satulovsky}. Thus, considering as an example $\alpha=0$, we can obtain from (\ref{n4}), considering (\ref{n2}) and (\ref{n5}),
\begin{eqnarray}\nonumber
\frac{d}{dt}\,P_{i}(0) & = & -\frac{\beta}{z}\,\sum_{\epsilon=1}^{z}[P_{i,i+\epsilon}(01) + P_{i,i+\epsilon}(02) + P_{i,i+\epsilon}(03)] \\ \label{n7}
&& + \frac{\gamma}{z}\,\sum_{\epsilon=1}^{z}P_{i,i+\epsilon}(20)
\end{eqnarray}

This is an exact expression for the single site correlation $P_{i}(0)$, that depends on the two-site correlations $P_{i,i+\epsilon}(01), P_{i,i+\epsilon}(02)$ and $P_{i,i+\epsilon}(03)$. The mean-field approximation we considered in our work consists in ignoring any demographic fluctuations, i.e., we take $P_{ij}(\alpha\,\beta) = P_i(\alpha)\,P_j(\beta)$. This is usually called one-site approximation \cite{satulovsky}. Taking this approximation in account, Eq. (\ref{n7}) can be rewritten as
\begin{eqnarray}\nonumber
\frac{d}{dt}\,P_{i}(0) & = & -\frac{\beta}{z}\,\sum_{\epsilon=1}^{z}[P_{i}(0)P_{i+\epsilon}(1) + P_{i}(0)P_{i+\epsilon}(2) + P_{i}(0)P_{i+\epsilon}(3)] \\ \label{n8}
&& + \frac{\gamma}{z}\sum_{\epsilon=1}^{z}P_{i}(2)P_{i+\epsilon}(0)
\end{eqnarray}

As a second approximation, we can search for spatially homogeneous solutions such that $P_{i}(\alpha) = P_j(\alpha)$ \cite{mario_book,satulovsky}. Considering this approximation, we have for (\ref{n8})
\begin{eqnarray}\label{n9}
\frac{d}{dt}\,P_{i}(0) & = & -\beta\,P_{i}(0)\,[P_{i}(1) + P_{i}(2) + P_{i}(3)] + \gamma\,P_{i}(2)\,P_{i}(0)
\end{eqnarray}
Identifying $P_i(0)=s$, $P_i(1)=m$, $P_i(2)=r$ and $P_i(3)=r_I$, Eq. (\ref{n9}) becomes
\begin{equation}
\frac{d}{dt}\,s(t) = -\beta\,s(t)\,m(t) - \beta\,s(t)[r(t)+r_I] + \gamma\,s(t)\,r(t)  
\end{equation}
that is Eq. (\ref{eq2}) of the text. Eqs. (\ref{eq3}) and (\ref{eq4}) can be obtained in an analogous way.


Now we can deal with the stationary solutions of Eqs. (\ref{eq2}) - (\ref{eq4}). Taking the limit $t\to\infty$ in Eq. (\ref{eq2}), we obtain $(-\beta\,m - \beta(r+r_{I})+\gamma\,r)\,s = 0$, where we denoted the stationary fractions as $s=s(t\to\infty), m=m(t\to\infty)$ and $r=r(t\to\infty)$. This last equation presents two solutions, namely $s=0$ and
\begin{equation} \label{eqx1}
r = \frac{\beta(m+r_{I})}{\gamma - \beta}
\end{equation}

If $s=0$ solution, the limit $t\to\infty$ in Eq. (\ref{eq3}) leads to a solution $m=0$. From the normalization condition, Eq. (\ref{eq1}), we found $r=1-r_I$. This set $(s,m,r)=(0,0,1-r_I)$ represents the absorbing phase found in section $3.1$ (See also Fig. \ref{fig2}).

On the other hand, if the valid solution is $s\neq 0$, Eq. (\ref{eqx1}) is valid, but we have also to consider the long-time limit in another equation. Taking the limit $t\to\infty$ in Eq. (\ref{eq3}), we have
\begin{equation} \label{eqx2}
\beta\,s\,(r+r_{I}) = [\alpha + \delta(r+r_{I}) - \beta\,s]\,m ~. 
\end{equation}

Considering Eqs. (\ref{eq1}) and (\ref{eqx1}), one obtains
\begin{equation}\label{app1}
s = 1 - \frac{\gamma\,(m+r_{I})}{\gamma - \beta}
\end{equation}

Substituting Eq. (\ref{eqx1}) in Eq. (\ref{eqx2}), we have
\begin{equation}\label{app2}
\beta\,s\left[\frac{\beta(m+r_{I})}{\gamma-\beta} + r_{I}\right] = \left\{\alpha + \delta\left[\frac{\beta(m+r_{I})}{\gamma-\beta} + r_{I}\right] -\beta\,s  \right\}\,m
\end{equation} 
For this last result, one can see that $m=0$ is only a solution for $r_{I}=0$, as discussed in \cite{nuno_lucas}. For $m\neq 0$, substituting Eq. (\ref{app1}) in (\ref{app2}), we find a second order polynomial for $m$ of the type $A\,m^{2} + B\,m + C=0$, where
\begin{eqnarray} \label{app3}
A & = & \frac{\gamma^{2}}{\gamma-\beta} + \delta \\ \label{app4}
B & = & r_I\left(\frac{\gamma(\gamma+\beta)}{\gamma-\beta}+\frac{\delta\gamma}{\beta}+\gamma\right) + \frac{\gamma\alpha}{\beta}-(\alpha+\gamma) \\ \label{app5}
C & = & \gamma\,r_I\left[\frac{\gamma\,r_I}{\gamma-\beta}-1\right]
\end{eqnarray}

This second order polynomial also indicates that $m=0$ is not a solution for the system with $r_I\neq 0$. Such polynomial gives us the stationary solution for $m$. Substituting the result in Eq. (\ref{app1}), we can find the stationary solution for $s$. Finally, the stationary solution for $r$ can be found from the normalization condition, Eq. (\ref{eq1}). This set $(s,m,r)$ represents the solution for the coexistence phase (see the discussion in section $3.1$ and Fig. \ref{fig2}).


\section{Analytical considerations: Model in the presence of $S_{I}$ inflexible individuals}

Taking the limit $t\to\infty$ in Eq. (\ref{eq10}), we obtain
\begin{equation}\label{app6}
(\alpha+\delta\,r)m = \gamma(s+s_I)r
\end{equation}

Considering the normalization condition (\ref{eq7}) written in the form $s+s_I=1-m-r$, Eq. (\ref{app6}) gives us
\begin{equation}\label{app7}
m = \frac{\gamma\,r\,(1-r)}{\alpha+(\gamma+\delta)\,r}
\end{equation}
Eq. (\ref{app7}) gives us a function $m=m(r)$.

Taking the limit $t\to\infty$ in Eq. (\ref{eq8}), and again considering the normalization $s+s_I=1-m-r$, we obtain
\begin{equation}\label{app8}
s = \frac{\gamma(1-m-r)r}{\beta(m+r)}
\end{equation}

Eq. (\ref{app8}) gives us a function $s=s(m,r)$. Substituting Eq. (\ref{app7}) in (\ref{app8}) and considering again the normalization, we obtain the following equation for $r$,
\begin{eqnarray}\label{app9}
r\,\{\gamma(\gamma\,r^{2}+\delta\,r+\alpha) -\gamma\,r\,[\alpha+(\delta+\gamma)\,r]-\beta\,(1-s_I)\,(\delta\,r+\alpha+\gamma)-C(r)\}=0 ~,
\end{eqnarray}
\noindent
where the function $C(r)$ is given by
\begin{eqnarray}\label{app10}
C(r)=\frac{\beta}{\alpha+(\delta+\gamma)\,r}\,[\delta^{2}\,r^{3}+2(\alpha+\gamma)\,\delta\,r^{2}+(\alpha+\gamma)^2\,r]
\end{eqnarray}

From Eq. (\ref{app9}), considering the function (\ref{app10}), we can see that $r=0$ is a solution (absence of noninflexible risk drinkers in the stationary states). If $r=0$ is a solution, from Eq. (\ref{app7}) we have that $m=0$ is also a solution. Thus, for $m=r=0$, the normalization condition Eq. (\ref{eq7}) gives us $s=1-s_{I}$. This set $(s,m,r)=(1-s_I,0,0)$ represents the absorbing phase found in section $3.2$ (See also Fig. \ref{fig4}).



\section*{References}


\begin{thebibliography}{00}



\bibitem{bailey1975mathematical}
N. T. Bailey, \textit{The mathematical theory of infectious diseases and its applications} (Charles Griffin \& Company Ltd, 5a Crendon Street, High Wycombe, Bucks HP13 6LE, 1975).

\bibitem{nuno_jorge}
N. Crokidakis, J. S. S\'a Martins, \textit{Can honesty survive in a corrupt parliament?}, Int. J. Mod. Phys. C 29, 1850094 (2018).

\bibitem{lima2014evolution}
F. Lima, T. Hadzibeganovic, D. Stauffer, \textit{Evolution of tag-based cooperation on Erd{\H{o}}s-R\'enyi random graphs}, International Journal of Modern Physics C 25 (2014) 1450006.

\bibitem{ejima2013modeling}
K. Ejima, K. Aihara, H. Nishiura, \textit{Modeling the obesity epidemic: social contagion and its implications for control}, Theoretical Biology and Medical Modelling 10 (2013) 17.

\bibitem{marvel2012encouraging}
S. A. Marvel, H. Hong, A. Papush, S. H. Strogatz, \textit{Encouraging moderation: clues from a simple model of ideological conflict}, Physical Review Letters 109 (2012) 118702.

\bibitem{stauffer2007can}
D. Stauffer, M. Sahimi, \textit{Can a few fanatics influence the opinion of a large  segment  of  a  society?}, The European Physical Journal B 57 (2007) 147-152.

\bibitem{daley1964epidemics}
D. J. Dailey, D. G. Kendall, \textit{Epidemics and Rumours}, Nature 204 (1964) 1118-1118.

\bibitem{gunduz2016dynamics}
G. G{\"u}nd{\"u}z, \textit{The dynamics of the rise and fall of empires},  International Journal of Modern Physics C 27 (2016) 1650123.

\bibitem{nizamani2014public}
S. Nizamani, N. Memon, S. Galam, \textit{From public  outrage  to  the  burst  of  public  violence:  An epidemic-like  model}, Physica A:  Statistical  Mechanics  and  its  Applications 416 (2014) 620-630.

\bibitem{galam2016modeling}
S. Galam, M. A. Javarone, \textit{Modeling radicalization phenomena in heterogeneous populations}, PloS one 11 (2016) e0155407.

\bibitem{brum2017dynamics}
R.  M.  Brum,  N.  Crokidakis, \textit{Dynamics  of  tax  evasion  through  an epidemic-like model}, International Journal of Modern Physics C 28 (2017) 1750023.

\bibitem{ijmpc2022}
N. Crokidakis, \textit{A simple mechanism leading to first-order phase transitions in a model of tax evasion}, International Journal of Modern Physics C 33, 2250075 (2022).

\bibitem{nuno_guns}
N. Crokidakis, \textit{Modeling the impact of civilian firearm ownership in the evolution of violent crimes}, Applied Mathematics and Computation 479, 127256 (2022). 

\bibitem{PMID:21696936}
F. Guerrero, F.-J. Santonja, R.-J. Villanueva, \textit{Analysing the Spanish smoke-free legislation of 2006: a new method to quantify its impact using a dynamic model}, The International Journal on Drug Policy 22 (2011) 247-251.

\bibitem{santonja2010alcohol}
F.-J. Santonja, E. S\'anchez, M. Rubio, J.-L. Morera,  \textit{Alcohol consumption in Spain and its economic cost: a mathematical modeling approach}, Mathematical and Computer Modelling 52 (2010) 999-1003.

\bibitem{sanchez2011predicting}
E. S{\'a}nchez, R.-J. Villanueva, F.-J. Santonja, M. Rubio, \textit{Predicting cocaine consumption in Spain: A mathematical modelling approach}, Drugs: Education, Prevention and Policy 18 (2011) 108-115.

\bibitem{heroine2021}
S. Djilali, S. Bentout, T.M. Touaoula, A. Tridane, S. Kumar, \textit{Global behavior of Heroin epidemic model with time distributed delay and nonlinear incidence function}, Results In Physics 31 (2021) 104953.

\bibitem{morrislarsen}
H. Morris, J. Larsen, E. Catterall, A.C. Moss, S.U. Dombrowski, \textit{Peer pressure and alcohol consumption in adults living in the UK: a systematic qualitative review}, BMC Public Health 20 (2020) 1014.

\bibitem{galea2009social}
S.  Galea,  C.  Hall,  G.  A.  Kaplan,  \textit{Social epidemiology  and complex system dynamic modelling as applied to health behaviour and drug use research}, International Journal of Drug Policy 20 (2009) 209-216.

\bibitem{gorman2006agent}
D.  M.  Gorman,  J.  Mezic,  I.  Mezic,  P. J. Gruenewald,  \textit{Agent-based modeling of drinking behavior: a preliminary model and potential applications to theory and practice},  American Journal of Public Health 96 (2006) 2055-2060.

\bibitem{sullivan2005}
L.  E.  Sullivan,  D.  A.  Fiellin,  P.  G.  O'Connor, \textit{The  prevalence  and impact  of alcohol  problems in  major  depression:  a  systematic  review}, The American Journal of Medicine 118 (2005) 330-341.

\bibitem{clay2020alcohol}
J. M. Clay, M. O. Parker,  \textit{Alcohol use and misuse during the COVID-19 pandemic: a potential public health crisis?}, The Lancet Public Health 5 (2020) e259.

\bibitem{narasimha2020complicated}
V. L. Narasimha, L. Shukla, D. Mukherjee, J. Menon, S. Huddar, U. K. Panda, J. Mahadevan, A. Kandasamy, P. K. Chand, V. Benegal, P. Murthy, \textit{Complicated alcohol withdrawal - An unintended consequence of COVID-19 lockdown}, Alcohol and Alcoholism (Oxford, Oxfordshire) (2020).

\bibitem{rehm2020alcohol}
J. Rehm, C. Kilian, C. Ferreira-Borges, D. Jernigan, M. Monteiro, C. D. Parry,  Z.  M.  Sanchez,  J.  Manthey,   \textit{Alcohol  use  in  times  of  the  COVID-19:  Implications for monitoring and policy}, Drug and Alcohol Review 39 (2020) 301-304.

\bibitem{nuno_lucas}
N. Crokidakis, L. Sigaud, \textit{Modeling the evolution of drinking behavior: A Statistical Physics perspective}, Physica A 570, 125814 (2021)

\bibitem{oms2018}
WHO, \textit{World Health Statistics 2018: Monitoring  health  for  the  sdgs, sustainable development goals}, Geneva:  World Health Organization CCBY-NC-SA (2018) 3.0 IGO.

\bibitem{galam1991}
S. Galam, S. Moscovici, \textit{Towards a theory of collective phenomena: Consensus and attitude changes in groups}, Eur. J. of Social Psychology 21, 49 (1991)

\bibitem{martins}
A. C. R. Martins, S. Galam, \textit{Building up of individual inflexibility in opinion dynamics}, Phys. Rev. E 87, 042807 (2013)

\bibitem{celia_vitor}
N. Crokidakis, V. H. Blanco, C. Anteneodo, \textit{Impact of contrarians and intransigents in a kinetic model of opinion dynamics}, Phys. Rev. E 89, 013310 (2014).

\bibitem{mobilia1}
M. Mobilia, A. Petersen, S. Redner, \textit{On the Role of Zealotry in the Voter Model}, J. Stat. Mech. P08029 (2007).

\bibitem{walters2013modelling}
C. E. Walters, B. Straughan, J. R. Kendal, \textit{Modelling alcohol problems: total recovery}, Ricerche di Matematica 62 (2013) 33-53.

\bibitem{nazir2019conformable}
A. Nazir, N. Ahmed, U. Khan, S. T. Mohyud-Din, \textit{A conformable mathematical model for alcohol consumption in Spain},  International Journal of Biomathematics 12 (2019) 1950057.

\bibitem{huo2018dynamics}
H.-F. Huo, H.-N. Xue, H. Xiang,  \textit{Dynamics of an alcoholism model on complex  networks  with  community  structure  and  voluntary  drinking}, Physica A: Statistical Mechanics and its Applications 505 (2018) 880-890.

\bibitem{mulone2012modeling}
G. Mulone, B. Straughan, \textit{Modeling binge drinking}, International Journal of Biomathematics 5 (2012) 1250005.

\bibitem{sanchez2007drinking}
F. S\'anchez, X. Wang, C. Castillo-Ch\'avez, D. M. Gorman, P. J. Gruenewald,  \textit{Drinking as an epidemic- a simple mathematical model with recovery and relapse}, in: Therapist’s Guide to Evidence-Based Relapse Prevention, Elsevier (2007) 353-368.

\bibitem{sharma2013drinking}
S. Sharma, G. P. Samanta, \textit{Drinking as an Epidemic: A mathematical model with dynamic behaviour},  Journal of Applied Mathematics \& Informatics 31 (2013) 1-25.

\bibitem{agrawal2018role}
A. Agrawal, A. Tenguria, G. Modi, \textit{Role of epidemic model  to  control  drinking  problem}, International Journal of Scientific Research in Mathematical and Statistical Sciences 5(4) (2018) 324-337.

\bibitem{adu2017mathematical}
I. K. Adu, M. AL-Rahman EL-Nor Osman, C. Yang, \textit{Mathematical model of drinking epidemic}, Journal of Advances in Mathematics and Computer Science, 22(5) (2017) 1-10. https://doi.org/10.9734/BJMCS/2017/33659.

\bibitem{huo2012global}
Hai-Feng Huo, Na-Na Song, \textit{Global stability for a binge drinking model with two stages}, Discrete Dynamics in Nature and Society, vol. 2012, Article ID 829386, 15 pages, 2012. https://doi.org/10.1155/2012/829386.

\bibitem{khajji2020discrete}
B. Khajji, A. Labzai, A. Kouidere, O. Balatif,  M. Rachik, \textit{A discrete mathematical modeling of the influence of alcohol treatment centers on the drinking dynamics  using  optimal control},  Journal of Applied Mathematics, vol. 2020, Article ID 9284698, 13 pages, 2020. https://doi.org/10.1155/2020/9284698.

\bibitem{huo2017optimal}
Hai-Fen Huo, Shui-Rong Huang, Xun-Yang Wang, H. Xiang, \textit{Optimal control of a social epidemic model with media coverage}, Journal of Biological Dynamics 11 (2017) 226-243.

\bibitem{muthuri2019modeling}
G. G. Muthuri, D. M. Malonza, F. Nyabadza,  \textit{Modeling the effects of treatment on alcohol abuse in Kenya incorporating mass media campaign}, Journal of Mathematical and Computational Science 9 (2019) 632-653.

\bibitem{ma2015modelling}
Shuang-Hong Ma, Hai-Feng Huo, Xin-You Meng, \textit{Modelling alcoholism as a contagious disease: A mathematical model with awareness programs and time delay}, Discrete Dynamics in Nature and Society 2015 (2015).

\bibitem{wang2014optimal}
Xun-Yang Wang, Hai-Feng Huo, Qing-Kai Kong, Wei-Xuan Shi, \textit{Optimal control strategies in an alcoholism model}, in:  Abstract and Applied Analysis, volume 2014, Hindawi, 2014.

\bibitem{din}
A. Din, Y. Li, \textit{The extinction and persistence of a stochastic model of drinking alcohol}, Results in Physics 28, 104649 (2021).

\bibitem{rahman}
M. ur Rahman, M. Arfan, Z. Shah, E. Alzahrani, \textit{Evolution of fractional mathematical model for drinking under Atangana-Baleanu Caputo derivatives}, Physica Scripta 96, 115203 (2021).

\bibitem{atman}
A. P. F. Atman, R. Dickman, J. G. Moreira, \textit{Phase diagram of a probabilistic cellular automaton with three-site interactions}, Phys. Rev. E 67, 016107 (2003).

\bibitem{galam_jacobs}
S. Galam, F. Jacobs, \textit{The role of inflexible minorities in the breaking of democratic opinion dynamics}, Physica A 381, 366 (2007).

\bibitem{galam2011}
S. Galam, \textit{Collective beliefs versus individual inflexibility: The unavoidable biases of a public debate}, 

\bibitem{mobilia2}
M. Mobilia, \textit{Nonlinear q-voter model with inflexible zealots}, Phys. Rev. E 92, 012803 (2015).

\bibitem{nuno_pmco}
N. Crokidakis, P. M. C. de Oliveira, \textit{Inflexibility and independence: Phase transitions in the majority-rule model}, Phys. Rev. E 92, 062122 (2015).


\bibitem{marro2005nonequilibrium}
J. Marro, R. Dickman, \textit{Nonequilibrium phase transitions in lattice models} (Cambridge University Press, 2005).

\bibitem{hinrichsen2000non}
H.  Hinrichsen,  Non-equilibrium  critical  phenomena  and  phase  transitions into absorbing states,  Advances in Physics 49 (2000) 815-958.

\bibitem{stanley_book}
H. E. Stanley, \textit{Introduction to Phase Transitions and Critical Phenomena} (Oxford University Press, 1987).



\bibitem{gomberg}
E.S.L. Gomberg, \textit{ALCOHOLIC WOMEN IN TREATMENT: THE QUESTION OF STIGMA AND AGE}, Alcohol \& Alcoholism 23, 507-514 (1988).

\bibitem{kolosova}
N.K. Rzhevskaya, V.A. Ruzhenkov, V.V. Ruzhenkova, U.S. Moskvitina, M.A. Kolosova, \textit{Coercion, Violation of Privacy and Everyday Difficulties as the Cause of Patient Refusal Treatment in Psychiatric Hospitals in Russia}, Int. J. of Criminology and Sociology 9, 968-973 (2020).

\bibitem{spirito}
A. Spirito, H. Sindelar-Manning, S.M. Colby, N.P. Barnett, W. Lewander, D.J. Rohsenow, P.M. Monti, \textit{Individual and family motivational interventions for alcohol-positive adolescents treated in an emergency department}, Arch. Pediatr. Adolesc. Med. 165, 269-274 (2011).

\bibitem{nsduh2019}
``Key Substance Use and Mental Health Indicators in the United States: Results from the 2019 National Survey on Drug Use and Health'', U.S. Department of Health and Human Services, 2020

\bibitem{mobilia3}
M. Mobilia, \textit{Commitment Versus Persuasion in the Three-Party Constrained Voter Model}, J. Stat. Phys. 151, 69-91 (2013).






\bibitem{belanger}
D. P. Belanger, \textit{Spin Glasses and Random Fields}, edited by A. P. Young (World Scientific, Singapore, 1998).
  
\bibitem{nuno_nobre}
N. Crokidakis, F. D. Nobre, \textit{Ising spin glass under continuous-distribution random magnetic fields: Tricritical points and instability lines}, Phys. Rev. E 77, 041124 (2008).
  
\bibitem{Barghathi1}
H. Barghathi, T. Vojta, \textit{Random Fields at a Nonequilibrium Phase Transition}, Phys. Rev. Lett. 109, 170603 (2012).

\bibitem{Barghathi2}
H. Barghathi, T. Vojta, \textit{Random field disorder at an absorbing state transition in one and two dimensions}, Phys. Rev. E 93, 022120 (2016).

\bibitem{holmes2019}
J. Holmes, A.K. Ally, P.S. Meier, R. Pryce, \textit{The collectivity of British alcohol consumption trends across different temporal processes: A quantile age–period–cohort analysis}, Addiction 114 (2019) 1970-1980.

\bibitem{conroy2013}
D. Conroy, R. de Visser, \textit{Being a non-drinking student: an interpretative phenomenological analysis}, Psychology \& Health 29 (2014) 536-551.



\bibitem{vankampen}
N. G. van Kampen, \textit{Stochastic Processes in Physics and Chemistry} (North-Holland, Amsterdan, 2007).

\bibitem{saeedian}
M. Saeedian, M. San Miguel, R. Toral, \textit{Absorbing phase transition in the coupled dynamics of node and link states in random networks}, Scientific Reports 9, 9726 (2019).

\bibitem{assaf}
M. Assaf and B. Meerson, \textit{WKB theory of large deviations in stochastic populations}, Journal of Physics A 50, 263001 (2017).




\bibitem{oliveira_quasi1}
M. M. de Oliveira, R. Dickman, \textit{Quasi-stationary simulation: the subcritical contact process}, Braz. J. Phys. 36, 685 - 689 (2006).

\bibitem{oliveira_quasi2}
M. M. de Oliveira, R. Dickman, \textit{How to simulate the quasistationary state}, Phys. Rev. E 71, 016129 (2005).






\bibitem{nuno_pmco2}
N. Crokidakis, P. M. C. de Oliveira, \textit{Impact of site dilution and agent diffusion on the critical behavior of the majority-vote model}, Phys. Rev. E 85, 041147 (2012).

\bibitem{manthey2008}
J. L. Manthey, A. Y. Aidoo, K. Y. Ward, \textit{Campus drinking: an epidemiological model}, J. Biol. Dynamics 2, 346-356 (2008).

\bibitem{galam_cont}
S. Galam, \textit{Minority opinion spreading in random geometry}, Eur. Phys. J. B 25, 403 (2002).

\bibitem{galam_cont2}
S. Galam, \textit{Contrarian deterministic effects on opinion dynamics: "the hung elections scenario"}, Physica A 333, 453 (2004).


\bibitem{nuno_joao}
J. P. Gambaro, N. Crokidakis, \textit{The influence of contrarians in the dynamics of opinion formation}, Physica A 486, 465 (2017).


\bibitem{stauffer_jorge}
D. Stauffer, J. S. S\'a Martins, \textit{Simulation of Galam's contrarian opinions on percolative lattices}, Physica 334, 558 (2004).




\bibitem{gutkin}
B. S. Gutkin, S. Dehaene , and J. -P. Changeux, \textit{A neurocomputational hypothesis for nicotine addiction}, Proceedings of the National Academy of Sciences 103, 1106-1111 (2006).


\bibitem{chou}
T. Chou, M. R. D'Orsogna, \textit{A mathematical model of reward-mediated learning in drug addiction}, Chaos 32, 021102 (2022).

\bibitem{guo2018}
Z.-K. Guo, H.-F. Huo, H. Xiang, \textit{Bifurcation analysis of an age-structured
alcoholism model}, J. Biol. Dynamics 12, 1009-1033 (2018).

\bibitem{liu2019}
L. Liu, X. Liu, \textit{Mathematical Analysis for an Age-Structured Heroin
Epidemic Model}, Acta Appl. Math. 164, 192-217 (2019).

\bibitem{chuang2018}
Y.-L. Chuang, T. Chou, M. R. D'Orsogna, \textit{Age-structured social interactions enhance
radicalization}, J. Math. Sociol. 42, 128-151 (2018).

\bibitem{mario_book}
T. Tom\'e, M. J. de Oliveira, \textit{Stochastic Dynamics and Irreversibility} (Springer, Heidelberg, 2015).


\bibitem{satulovsky}
J. E. Satulovsky, T. Tom\'e, \textit{Stochastic lattice gas model for a predator-prey system}, Phys. Rev. E 49, 5073 (1994).




\end{thebibliography}
\end{document}